\documentclass[conference]{IEEEtran}
\IEEEoverridecommandlockouts

\usepackage{cite}
\usepackage{amsmath,amssymb,amsfonts}
\usepackage{algorithmic}
\usepackage{graphicx}
\usepackage{textcomp}
\usepackage{xcolor}
\usepackage{times}  
\usepackage{mathptmx}  
\usepackage{newtxtext,newtxmath}  
\def\BibTeX{{\rm B\kern-.05em{\sc i\kern-.025em b}\kern-.08em
    T\kern-.1667em\lower.7ex\hbox{E}\kern-.125emX}}
\begin{document}

\newcommand{\todo}[1]{{\color{red}#1}}
\newcommand{\done}[1]{{\color{blue}#1}}


\title{Measurement-Based Line-Impedance Estimation in the Absence of Phasor Measurement Units
\thanks{This work was supported by HES-SO under the projects RESINET and EXPOSITION.}
}

\author{\IEEEauthorblockN{Plouton Grammatikos}
\IEEEauthorblockA{\textit{School of Engineering} \\
\textit{HES-SO Valais-Wallis}\\
Sion, Switzerland \\
plouton.grammatikos@hevs.ch}
\and
\IEEEauthorblockN{Ali Mohamed Ali}
\IEEEauthorblockA{\textit{School of Engineering} \\
\textit{HES-SO Valais-Wallis}\\
Sion, Switzerland \\
ali.alimohamed@hevs.ch}
\and
\IEEEauthorblockN{Fabrizio Sossan}
\IEEEauthorblockA{\textit{School of Engineering} \\
\textit{HES-SO Valais-Wallis}\\
Sion, Switzerland \\
fabrizio.sossan@hevs.ch}
}

\maketitle

\begin{abstract}

This paper proposes and compares experimentally several methods to estimate the series resistance and reactance (i.e., the transversal components of the $\pi$-model of a line) of low-voltage lines in distribution grids. It first shows that if phasor measurements are available and the grid nodal voltages and power injections are known, the problem can be formulated and solved as a conventional load flow with properly adjusted unknowns. To solve this problem, we propose an analytical derivation of the Jacobian matrix. If only RMS values are available, such as from smart meters, integrating information from multiple intervals becomes necessary, ultimately opening to least-squares estimations, widely adopted in the literature. In this context, applying the proposed Jacobian contributes to accelerating the problem resolution of existing algorithms. The methods are compared in terms of estimation performance and convergence by using measurements from an experimental distribution grid interfacing real-world components and with realistic size implemented at the Gridlab at HES-SO Valais.
\end{abstract}


\section{Introduction}
Increasing distributed generation, particularly photovoltaic (PV), and electrification trends, such as heat pumps and electric vehicles (EVs), challenge the power distribution infrastructure. Accommodating these needs might require costly grid reinforcement to ensure that the line and transformer currents and nodal voltage magnitudes stay within physical and statutory limits (e.g., \cite{crozier2024distribution, gupta2021countrywide, steinbach2024grid}). Leveraging flexible resources within the distribution grid to restore appropriate operations stems as an alternative to grid reinforcement, which is widely advocated in the literature. 
Flexible resources include dedicated assets operated by the distribution system operator (DSOs), such as on-load tap-changer transformers and battery energy storage systems (BESSs), and behind-the-meter assets, such as curtailable PV, BESS in building premises, demand response, and EV uni- and bi-directional chargers.  Control strategies for flexible resources can be broadly classified as model-based and model-free. Model-based approaches use a grid model to compute the grid remedial actions to control flexible resources. They are typically formulated as an optimal power flow (OPF) where controllable resources are modeled as power injections subject to flexibility constraints (e.g., power and energy levels), and grid constraints are expressed with load flow equations, either linearized or relaxed, to retain problem traceability. 
In model-free approaches, a grid model is not available, and control decisions are computed based on local measurements, such as in droop and volt-var control. 

Model-based approaches critically depend on the knowledge of the grid model and the nodal power injections. The grid model, the focus of this paper, is usually estimated from topological information and cable electrical parameters from the DSO's records. While the grid topology can usually be inferred from the states of the breakers and fuses, line parameters might be subject to uncertainty or variations from recorded values. Sources of uncertainty include variable weather conditions (e.g., conductor, air and soil temperatures, humidity), cable state (e.g., insulation aging), non-ideal contact resistance in junction boxes, and record errors. Erroneous impedance information degrades load flow estimation performance and, in OPFs, might result in unfeasible control actions.

Reliable estimates of the line electrical parameters are thus critical for grid state-estimation procedures.
Estimating a line model entails identifying the values of the transversal components and shunt elements of the $\pi$-model of a transmission line; proposed estimators use either time- or phasor-domain information. Works in \cite{535680} and \cite{priyanka2023data} apply Kalman filtering and Gaussian-processes regression, respectively, to time-domain information. In the phasor domain, the work in \cite{7285678} uses phasor measurement unit (PMU) measurements and Kalman filtering to estimate the parameters of a medium-length transmission line. The work in \cite{9464332} applies PMU measurements and total-least squares (LS) to estimate the admittance matrix of the grid. The work in \cite{wang2021estimate} applies graphical learning to PMU measurements. The work in \cite{9566736} proposes a two-step on-line process to identify first poorly modeled lines and then refine parameters numerically. The work in \cite{leal2024backward} combines LS and the inverse load flow problem. 

This paper first formulates the line parameter estimation problem by reframing the classical load flow analysis, providing a different insight than the literature. We show that if measurements from PMUs are available and the grid is fully observable, conventional Newton-Rhapson (NR) can be applied to estimate the lines' resistance and reactance. To this end and as a main methodological contribution, this paper develops analytically the Jacobian to solve this problem. If only RMS measurements are available instead (e.g., from a smart meter), solving the problem requires combining measurements from multiple time intervals. By combining even more measurements, we show that the problem can be formulated in the least-squares (LS) sense, establishing a link with the existing literature. In this case, we show that the proposed Jacobian helps accelerate the resolution process of off-the-shelf nonlinear LS using the trust-region-reflective (TRR) algorithm. The performance of all these methods is validated and compared experimentally in a low-voltage (LV) distribution grid of realistic size and with real-world components implemented in a laboratory setting. Finally, considerations of the convergence of these algorithms complete this work's contributions.

The rest of this paper is organized as follows: Section~\ref{sec:problemform} formulates the problem of line-parameter estimation, Section~\ref{sec:methods} describes the methods we propose to estimate the line parameters in different settings, Section~\ref{sec:exp_setup} validates the proposed methods through simulations and experiments, and Section~\ref{sec:concs} concludes the paper.

\section{Problem Formulation}\label{sec:problemform}
We consider a radial single-phase grid with $N$ nodes and $L=N-1$ lines (the three-phase case could be studied in future research). We denote by $Z_l=R_l+jX_l$ the impedance across line $l$, where $R_l$ is the resistance and $X_l$ is the reactance of the line. Line shunt elements are ignored under the assumption that low-voltage lines are short. The nodal admittance matrix of the grid is
\begin{equation}
    \label{eq:Y-matrix}
    Y = A^TY_{pr}A \in \mathbb{C}^{N \times N}
\end{equation}
where $Y_{pr} \in \mathbb{C}^{L \times L}$,  the primitive admittance matrix, is a diagonal matrix with elements
\begin{equation}
    \label{eq:primitive-Y-matrix}
    Y_{pr, l} = \frac{1}{Z_l},
\end{equation}
and $A \in \{0,1\}^{L\times N}$ is the branch-to-node matrix with elements
\begin{equation}
    a_{l, k}=
    \begin{cases}
        1 & \text{if node $k$ is at the top of line $l$} \\
        -1 & \text{if node $k$ is at the bottom of line $l$} \\
        0 & \text{if node $k$ is not connected to line $l$}.
    \end{cases}
\end{equation}

We also consider the power flow equations for all nodes $k=1...N$:
\begin{equation}
    \label{eq:power-flow-P}
    P_k = \sum_{j=1}^N |V_k||V_j||Y_{kj}|cos(\theta_k - \theta_j -\phi_{kj})
\end{equation}
\begin{equation}
    \label{eq:power-flow-Q}
    Q_k = \sum_{j=1}^N |V_k||V_j||Y_{kj}|sin(\theta_k - \theta_j -\phi_{kj}),
\end{equation}
where $P_k, Q_k$ are the active and reactive power injection at node $k$, $V_k = |V_k| \angle \theta_k$ is the voltage phasor of node $k$ and $Y_{kj} = |Y_{kj}| \angle \phi_{kj}$ are the elements of the nodal admittance matrix.
Given some initial estimates $\hat{R}_{l,0}$ and $\hat{X}_{l,0}$ of the line resistances and reactances, our goal is to compute the improved estimates $\hat{R}_l$ and $\hat{X}_l$.

\section{Estimating the Line Parameters}\label{sec:methods}

\subsection{The Newton-Raphson Method}
\label{sec:newton-raphson}

We present two methods to correct the line parameter estimates based on NR. NR finds the solution of $K$ continuously differentiable nonlinear equations $f_k(\mathbf{x})=0, k=1, \dots, K$ in $K$ unknowns $\mathbf{x}=[x_1,x_2,\dots,x_K]^T$. This is equivalent to finding the roots of the vector function $\mathbf{F}(\mathbf{x})=[f_1(\mathbf{x}),f_2(\mathbf{x}),\dots,f_K(\mathbf{x})]^T$. At each iteration $n$, NR computes the difference $\Delta \mathbf{x}_n$ by solving the system of linear equations
\begin{equation}
    \label{eq:nr-linear-system}
    \mathbf{F(\mathbf{x}_n)} = -J_{\mathbf{F}}(\mathbf{x}_n) \Delta \mathbf{x}_n
\end{equation}
where $\mathbf{x}_n$ is the estimation of $\mathbf{x}$ at iteration $n$ and $J_{\mathbf{F}}$ is the Jacobian matrix of $\mathbf{F}$ with elements as:
\begin{equation}
    J_{\mathbf{F}}(\mathbf{x})_{ij} = \frac{\partial f_i}{\partial x_j}(\mathbf{x}), i,j \in \{1,2,\dots,K\}.
\end{equation}
Then, the new estimation of $\mathbf{x}$ is given by
\begin{equation}
    \label{eq:nr-step}
    \mathbf{x}_{n+1} = \mathbf{x}_n + \alpha \Delta \mathbf{x}_n
\end{equation}
where $\alpha \leq 1$ is an adjustable step size that prevents the solution estimate from oscillating. The method is repeated until the value of $||\Delta 
 \mathbf{x}_n||_\infty$ is below a given tolerance.

\subsection{First Method: Using PMUs}
\label{sec:with-pmus}

We assume a radial grid where the voltage phasors $V_k$ and complex power injections $P_k + j Q_k$ are known from measurements at all nodes $k=1,\dots, N$.

Writing expressions \eqref{eq:power-flow-P}-\eqref{eq:power-flow-Q} results in $2N$ equations and $2(N-1)=2N-2$ unknowns, which we collect in the vector $\mathbf{x}=[R_1, \dots, R_L, X_1, \dots, X_L]^T$. In this setting and by dropping the measurements for one node, it is possible to solve the equations in the problem unknowns by applying NR.

By arbitrarily dropping information from node 1, assumed to be the slack bus, the vector function $\mathbf{F}$ is
\begin{equation}
    \mathbf{F} = [\Delta P_2, \dots, \Delta P_N, \Delta Q_2, \dots, \Delta Q_N]^T,
\end{equation}
where
\begin{equation}
    \label{eq:delta_P}
 \Delta P_k = -P_k + \sum_{j=1}^N |V_k||V_j||Y_{kj}|cos(\theta_k - \theta_j -\phi_{kj})
\end{equation}
\begin{equation}
    \label{eq:delta_Q}
    \Delta Q_k = -Q_k + \sum_{j=1}^N |V_k||V_j||Y_{kj}|sin(\theta_k - \theta_j -\phi_{kj}).
\end{equation}

Computing the Jacobian matrix $J_{\mathbf{F}}$ requires the derivatives of $\Delta P_k$ and $\Delta Q_k$ with respect to $R_l$ and $X_l$. These can be computed analytically as follows (and similarly for $X_l$):
\begin{align}
    \begin{aligned}
        \frac{\partial \Delta P_k}{\partial R_l} &= \sum_{j=1}^N |V_k||V_j|\frac{\partial |Y_{kj}|}{\partial R_l}cos(\theta_k-\theta_j-\phi_{kj})+\\
        &+|V_k||V_j||Y_{kj}|sin(\theta_k-\theta_j-\phi_{kj})\frac{\partial \phi_{kj}}{\partial R_l}
    \end{aligned} \label{eq:dPdR} \\
    \begin{aligned}
        \frac{\partial \Delta Q_k}{\partial R_l} &= \sum_{j=1}^N |V_k||V_j|\frac{\partial |Y_{kj}|}{\partial R_l}sin(\theta_k-\theta_j-\phi_{kj})+\\
        &-|V_k||V_j||Y_{kj}|cos(\theta_k-\theta_j-\phi_{kj})\frac{\partial \phi_{kj}}{\partial R_l}.
    \end{aligned} \label{eq:dQdR}
\end{align}

We now compute the partial derivatives of $|Y_{kj}|$ with respect to $R_l$ and $X_l$, and of $\phi_{kj}$ with respect to $R_l$ and $Q_l$. From \eqref{eq:Y-matrix} and \eqref{eq:primitive-Y-matrix}, we get the elements of the nodal admittance matrix:
\begin{equation}
    \label{eq:Y-matrix-elements}
    Y_{kj} = \sum_{l=1}^L a_{lk}a_{lj} \frac{1}{R_l+jX_l} = \sum_{l=1}^L a_{lk}a_{lj} \frac{R_l-jX_l}{R_l^2+X_l^2}.
\end{equation}
We then compute the derivative
\begin{equation}
    \label{eq:Y-complex-derivative}
    \frac{\partial Y_{kj}}{\partial R_l} = a_{lk}a_{lj}\frac{X_l^2-R_l^2+j2X_lR_l}{(R_l^2+X_l^2)^2}.
\end{equation}
Given the following definitions
\begin{equation}
    |Y_{kj}|=\sqrt{\Re(Y_{kj})^2+\Im(Y_{kj})^2},
\end{equation}
\begin{equation}
    \phi_{kj}=\arctan\left(\frac{\Im(Y_{kj})}{\Re(Y_{kj})}\right),
\end{equation}
where $\Re(\cdot)$ and $\Im(\cdot)$ denote the real and imaginary parts respectively, we can compute the following derivatives
\begin{equation}
    \frac{\partial |Y_{kj}|}{\partial R_l} = \frac{1}{|Y_{kj}|}\left(\Re(Y_{kj}) \frac{\Re(Y_{kj})}{\partial R_l}+\Im(Y_{kj})\frac{\partial \Im(Y_{kj})}{\partial R_l}\right)
\end{equation}
\begin{equation}
    \frac{\partial \phi_{kj}}{\partial R_l} = \frac{1}{1+\tan^2(\phi_{kj})} \frac{\Re(Y_{kj})\frac{\partial \Im(Y_{kj})}{\partial R_l}-\Im(Y_{kj})\frac{\partial \Re(Y_{kj})}{\partial R_l}}{(\Re(Y_{kj}))^2}
\end{equation}
Finally, by linearity of the $\Re$ and $\Im$ operators, we have
\begin{equation}
    \begin{aligned}
        \frac{\partial \Re(Y_{kj})}{\partial R_l} = \Re\left( \frac{\partial Y_{kj}}{\partial R_l} \right), 
        \frac{\partial \Im(Y_{kj})}{\partial R_l} = \Im\left(\frac{\partial Y_{kj}}{\partial R_l}\right),
    \end{aligned}
\end{equation}
which combined with \eqref{eq:Y-complex-derivative} allows us to compute the desired derivatives. 
A similar formula can be derived in the same way for the derivatives with respect to $X_l$ by noting from \eqref{eq:Y-matrix-elements} that
$\frac{\partial Y_{kj}}{\partial X_l} = j\frac{\partial Y_{kj}}{\partial R_l}, \forall j,k,l$.

Note that for \eqref{eq:nr-linear-system} to have a unique solution, $J_F$ must not be singular; to this end, equations \eqref{eq:power-flow-P}-\eqref{eq:power-flow-Q} should be linearly independent. This can be ensured by choosing sufficiently large injections $P_k$ and $Q_k$ to exploit the non-linear characteristics of the power flow equations.

\subsection{Second Method: Without Using PMUs}
\label{sec:without-pmus}

If only the RMS values of the nodal voltages are available, then the number of unknowns in equations  \eqref{eq:power-flow-P}-\eqref{eq:power-flow-Q} increases by $N-1$, one for each unknown voltage angle (the voltage angle at the slack bus is fixed). In this setting, the total unknowns are $2(N-1)+N-1=3N-3$ for $2N$ equations; hence, for $N>3$, there are more unknowns than equations, and the system \eqref{eq:nr-linear-system} does not have a unique solution.

With two sets of power injections from two different time instances $t_1, t_2$, labeled as $\{P_k^{t_1}, Q_k^{t_1}, k=1, \dots, N\}$ and $\{P_k^{t_2}, Q_k^{t_2}, k=1, \dots, N\}$, expressions \eqref{eq:power-flow-P}-\eqref{eq:power-flow-Q} result in $4N$ equations and $2(N-1)$ unknowns, given by the line parameters and the $2(N-1)$ voltage angles for $t_1$ and $t_2$. Hence, the total number of unknowns is $4N-4$, smaller than the number of equations for every value of $N$. Therefore, two time steps should carry enough information to compute the line parameters for an arbitrarily large grid.

The unknown vector and the function $F$ are respectively
\begin{equation}
    \begin{aligned}
        \mathbf{x}=[&R_1, \dots, R_L, X_1, \dots, X_L,\\
        &\theta_2^{t_1}, \dots, \theta_N^{t_1}, \theta_2^{t_2}, \dots, \theta_N^{t_2}]^T
    \end{aligned}
\end{equation}
\begin{equation}
    \label{eq:F-function}
    \begin{aligned}
        \mathbf{F}=[&\Delta P_2^{t_1}, \dots, \Delta P_N^{t_1}, \Delta Q_2^{t_1}, \dots, \Delta Q_N^{t_1},\\
        &\Delta P_2^{t_2}, \dots, \Delta P_N^{t_2}, \Delta Q_2^{t_2}, \dots, \Delta Q_N^{t_2}]^T,
    \end{aligned}
\end{equation}
where $\Delta P_k^{t_i}$ and $\Delta Q_k^{t_i}, i\in\{1,2\}$ are given by \eqref{eq:delta_P} and \eqref{eq:delta_Q}. Again, we have arbitrarily ignored the power flows at the slack bus because there are more equations than unknowns. To compute all the elements of the new Jacobian, we also need to compute the derivatives $\frac{\partial \Delta P_k^{t_i}}{\partial \theta_k^{t_j}}$ and $\frac{\partial \Delta P_k^{t_i}}{\partial \theta_k^{t_j}}$ for $i,j\in\{1,2\}$. For $t_i=t_j$, the derivatives can easily be computed from \eqref{eq:delta_P} and \eqref{eq:delta_Q}. For $t_i \neq t_j$, the derivatives are 0, as the power flows in the two time instances are independent.

To minimize the risk of $J_F$ being singular, we choose $t_1, t_2$ so that the nodal power injections are as different as possible from each other; in this way, we would exploit the non-linear characteristics of the power flow equations to ensure that there is no linear dependence of the equations between the two time instances. The singularity of $J_F$ is studied in Section~\ref{sec:jacobian-singularity}.

\subsection{Least Squares Estimation}
\label{sec:least-squares}

In a real-life application, the voltage and power measurements might not satisfy the power-flow equations due to measurement errors and noise; this will negatively impact the NR, which might not converge to the correct value of the line parameters. To mitigate the effect of the noise, one could use LS estimation. We compare two methods to solve this problem.

\subsubsection{Newton-Raphson with an overdetermined system}

Sections~\ref{sec:with-pmus} and \ref{sec:without-pmus} showed that voltage phasors and nodal power injections are sufficient to estimate the line parameters, whereas, if phase angles are not available, information from two time intervals is required. 
However, if time series measurements of the relevant quantities are available, one could compute a least-sqaures estimation of the roots of $\mathbf{F}(\mathbf{x})$ to edge against measurement errors and noise.
With additional measurements, the system of equations \eqref{eq:delta_P} and \eqref{eq:delta_Q} becomes overdetermined, and $J_\mathbf{F}$ will not be square. In this case, we can approximate a solution of the system in \eqref{eq:nr-linear-system} by minimizing
\begin{equation}
    \min_{\Delta \mathbf{x}_n}||\mathbf{F(\mathbf{x}_n)} + J_{\mathbf{F}}(\mathbf{x}_n) \Delta \mathbf{x}_n||_2
\end{equation}
at each iteration of NR (see, for example, \cite{Barata_2011}). 

\subsubsection{Quadratic optimization}

Instead of NR, the estimation problem can be formulated as a constrained optimization problem. Given voltage and power measurements from $T$ timesteps, the problem is to minimize the following objective under constraints:
\begin{equation}
    \label{eq:least-squares}
    \begin{aligned}
        \min_{R_l,X_l,\theta_k^{t_i}} \sum_{i=1}^T\sum_{k=1}^N (\Delta P_k^{t_i})^2+(\Delta Q_k^{t_i})^2 \\
        \text{s.t. } R_l, X_l > 0, \forall l \in \{1,\dots,L\} \\
        \theta_k^{t_i} \in [-\pi, \pi], \forall k \in \{1,\dots,N\}, \forall t_i \in \{t_1,\dots,t_T\} \\
        \theta_1^{t_i} = 0, \forall t_i \in \{t_1,\dots,t_T\}
    \end{aligned}
\end{equation}
where the voltage angle at the slack bus (node 1) is customarily assumed to be 0 at all time intervals. It should be noted that the problem can be rewritten for the scenario of Section~\ref{sec:with-pmus} by using the values of the voltage angles given by the PMUs.
The problem in \eqref{eq:least-squares} can be solved with conventional quadratic programming techniques. In Section~\ref{sec:experiments}, we use the trust-region-reflective (TRR) algorithm \cite{doi:10.1137/0806023}, implemented by Matlab, and compare its performance to NR. The algorithm can also use the analytical computation of the Jacobian matrix to help with convergence. Finally, both LS methods require an initial impedance estimation, e.g, from the cable datasheet and length estimate. However, as shown in Section~\ref{sec:experiments}, both methods can converge to the correct impedance values even with rough initial estimates (e.g., up to a factor of 5).

\section{Experimental Validation}

\subsection{Experimental Setup}
\label{sec:exp_setup}

\begin{figure}[]
    \centering
    \includegraphics[width=\linewidth]{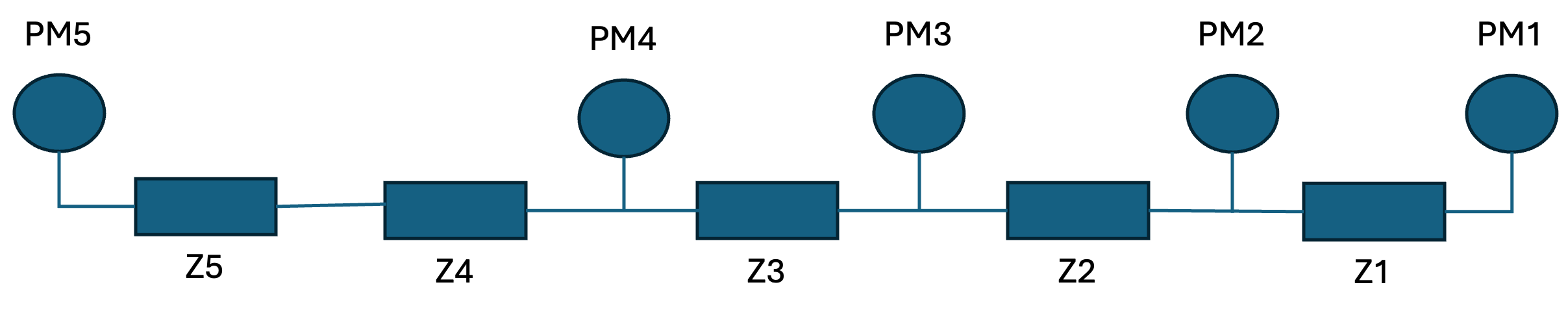}
    \caption{Schematic of district}
    \label{fig:district-schematic}
\end{figure}

The proposed method is experimentally tested and preliminarily validated on a distribution grid test bench of the GridLab laboratory at HES-SO Valais-Wallis in Sion. The platform is designed to emulate an LV distribution grid on a realistic scale and has 4 three-phase distribution lines called districts. The schematic of a single district is shown in Fig.~\ref{fig:district-schematic}. Each district has five nodes, labeled PM1-PM5. Node 5 serves as a connection point for external sources or loads of a maximum of 16~A per phase. Nodes 1-3 interface 4-quadrant load emulators with a rated power of 15~kVA (one per node), implemented by back-to-back AC/DC converter powered by a dedicated and separated supply. 
Node 4 can be used to connect external loads to the district. The active and reactive power injections, the RMS of all nodal voltages, and line currents are measured using Siemens SICAM P devices. The current sensors are class 1 at a rated current of 50~A.

Impedances $Z_1$-$Z_4$ are emulated to model 500-m-long aerial lines; $Z_5$ is a real 60~m cable. Table~\ref{tab:line_parameters} shows the line parameters per phase given by the manufacturer's datasheet. We aim to correct the estimation of these parameters using the proposed methods. Because in our setup, injecting power into PM4 is not possible, impedances $Z_3, Z_4$, and $Z_5$ are considered one.

\begin{table}[]
    \caption{Line parameters per phase given by datasheet}
    \centering
    \begin{tabular}{c|c|c}
         Line & Resistance ($m\Omega$) & Reactance ($m\Omega$)  \\
         \hline
         $Z_1, Z_2,Z_3,Z_4$ & 150 & 141.4\\
         $Z_5$ & 184.8 & 5.3
    \end{tabular}
    \label{tab:line_parameters}
\end{table}

\subsection{Preliminary validation with simulations}
\label{sec:simulations}

We first verify by simulations whether the proposed method can correct the estimation error of the line parameters. To this end, we generate the necessary measurements by a load flow where the true line parameters are perturbed randomly (uniform distribution) by $\pm 25\%$ with respect to the datasheet value. Starting from the initial estimation in the datasheet, we expect the algorithm to converge to the true value of the line parameters. Because PMU measurements are generally not available in LV distribution grids, we use the method of Section~\ref{sec:without-pmus} that relies on RMS measurements (e.g., from smart meters) at all nodes at two time instances:
\begin{equation}
    \label{eq:power_input}
    \begin{aligned}
        P^{t_1}_k=3000~\text{W}, Q^{t_1}_k=3000~\text{Var}, \forall k \in \{1,2,3\} \\
        P^{t_2}_k=-3000~\text{W}, Q^{t_2}_k=-3000~\text{Var}, \forall k \in \{1,2,3\}.
    \end{aligned}
\end{equation}
At this stage, the grid is modeled by way of a single-phase equivalent under the assumption that loads are balanced and that line impedances are the same. 

The voltage angles are computed by NR, as explained in Section~\ref{sec:without-pmus}. However, NR requires a reasonable initial estimate of the unknown vector $\mathbf{x}$ to guarantee convergence. 
For this purpose, we assume that the line impedances mainly affect the voltage magnitudes, not the angles. Therefore, we assume that a reasonable initial estimate for the voltage angles is given by load flow using the line parameters given by the datasheet. It should be noted that the power levels of \eqref{eq:power_input} were chosen such that the voltage magnitudes stay within a reasonable range from the nominal value ($\pm 10\%$).

The simulations are performed in Matlab. A tolerance value of $10^{-6}~\text{pu}$ for the convergence of the norm of $\Delta \mathbf{x}$ in \eqref{eq:F-function} is used. NR converged in five iterations using a step size of $\alpha=1$ in \eqref{eq:nr-step}. It was empirically observed that smaller values affected only the speed of convergence of NR and not the final value. 
Table~\ref{tab:estimation-error} shows the error between the estimation of the nodal power injections, the voltage phasors, and the line parameters and the respective true values before ($n=0$) and after the five iterations ($n=5$). We see that the percentage error of the estimation of the line parameters improves by two orders of magnitude. The algorithm also correctly estimates the voltage angle values for both time steps, with an error of $10^{-6}$~rad, denoting a good ability to retrieve the voltage phasor even without PMUs.

\begin{table}[]
    \caption{Estimation error before and after Newton-Raphson has converged}
    \centering
    \begin{tabular}{c|c|c}
    Value & Max estimation error $n=0$ & Max estimation error $n=5$ \\
    \hline
    $P^{t_1}$ & 561 W & $1.1 \times 10^{-4}$ W \\
    $P^{t_2}$ & 575 W & $1.4 \times 10^{-4}$ W \\
    $Q^{t_1}$ & 530 Var & $1.1 \times 10^{-4}$ Var \\
    $Q^{t_2}$ & 529 Var & $1.1 \times 10^{-4}$ Var\\
    $V^{t_1}$ & $0.0029$ pu & $8.7 \times 10^{-9}$ pu \\
    $V^{t_2}$ & $0.0034$ pu & $1.0 \times 10^{-10}$ pu\\
    $\theta^{t_1}$ & $0.0014$ rad & $1.9 \times 10^{-6}$ rad\\
    $\theta^{t_2}$ & $0.0015$ rad & $2.4 \times 10^{-6}$ rad\\
    $R_l$ & 25 \% & 0.10 \%\\
    $X_l$ & 19.5 \% & 0.11 \%\\
    \end{tabular}
    \label{tab:estimation-error}
\end{table}

\subsection{Singularity of Jacobian Matrix}
\label{sec:jacobian-singularity}

\begin{figure} []
    \centering
    \includegraphics[width=\linewidth]{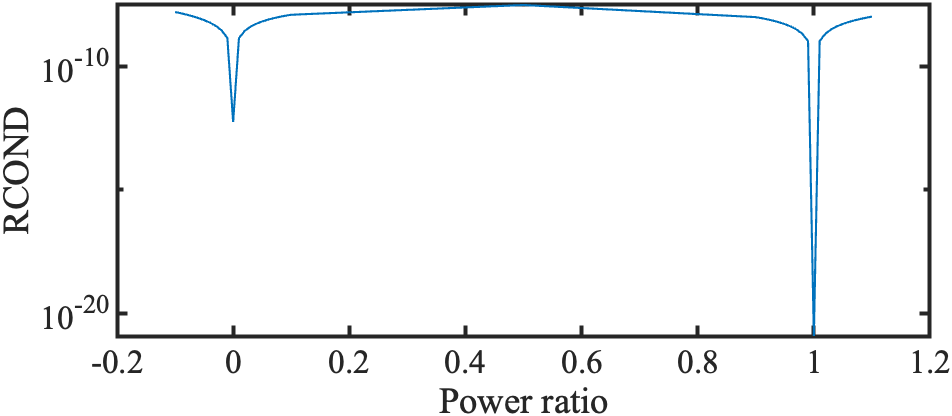}
    \caption{Reciprocal condition number (RCOND) of the Jacobian matrix as a function of the ratio between the power injections of the two time instances.}
    \label{fig:rcond}
\end{figure}

We now analyze the impact of measurement selections on NR's convergence. As explained in Section~\ref{sec:without-pmus}, nodal power injections at $t_1$ and $t_2$ should be such that the Jacobian matrix is non-singular. To characterize the solution's quality, we use the Jacobian's reciprocal condition number (RCOND), for which small values denote unreliable solutions.
We set the complex power injections at $t_1$ to 3~kW$+j$3~kVar; we then vary the power injections at $t_2$ as
\begin{equation}
    \label{eq:two-times}
    P^{t_2}_k=rP_k^{t_1}, Q^{t_2}_k=rQ_k^{t_1}, \forall k
\end{equation}
where coefficient $r \in [-0.1, 1.1]$ is called power ratio. The initial estimates of the voltage angles and the line parameters are the same as in Section~\ref{sec:simulations}. We compute the RCOND of the Jacobian for the first NR iteration. The results are shown in Fig.~\ref{fig:rcond}: the Jacobian becomes singular either when the powers at $t_1$ and $t_2$ are similar (less than 1\% difference) or when the powers at $t_2$ are close to 0 (by symmetry of \eqref{eq:two-times}, the same result would apply if the powers at $t_1$ were 0).

Therefore, as long as the power injections at the two-time instances are large enough and are not too similar (even 1\% difference is enough), the system \eqref{eq:nr-linear-system} will have a unique solution, and NR will converge. However, this does not guarantee that NR convergences to the line parameters' true value, as multiple values might satisfy the power flow equations for the given power injections. Therefore, we might need more measurements to get a more confident estimation. This is studied more in Section~\ref{sec:experiments}.

\subsection{Validation in an Experimental Setting}
\label{sec:experiments}

This section compares the line parameter estimation performance of NR and quadratic optimization with TRR, described in Section~\ref{sec:least-squares}, with experimental measurements. LS requires measurements at multiple time intervals. We reproduce in the experimental test-bench different combinations of active and reactive power at nodes 1-3 in Fig.~\ref{fig:district-schematic}. In particular, we inject all active and reactive power combinations in the range [-4, 4] kW and kVar with a step of 1~kW and kVar, respectively, to all nodes.
For each combination, measurements are averaged over a 1-minute interval to reduce measurement noise.

The computations are performed in MATLAB, using a custom NR implementation and the standard \emph{lsqnonlin} function for TRR; the latter accepts the Jacobian of the objective function, which we compute analytically in Section~\ref{sec:with-pmus}, as an optional argument to accelerate convergence.
The convergence tolerance for $\Delta \mathbf{x}$ is set to $10^{-6}$ in both algorithms.
For NR, the step size $\alpha$ is 0.1, as the previous value 1 caused oscillations.
Table~\ref{tab:line-parameters-estimation} compares the values of the complex line impedances estimated by NR and TRR against the original datasheet estimates:
both algorithms converge to approximately the same values; estimated $R$ and $X$ values are up to 67\% and 22\%, respectively, larger than datasheet information.

\begin{table}[]
    \caption{Estimated line parameters with Newton-Raphson and Trust-region-reflective algorithm}
    \centering
    \begin{tabular}{c|c|c|c}
    Method & $Z_1$ ($m\Omega$) & $Z_2$ ($m\Omega$) & $Z_{3-5}$ ($m\Omega$) \\
    \hline
    Datasheet & 150+$j$141.4 & 150+$j$141.4 & 484.8+$j$288.2 \\
    Newton-Raphson & 250.9+$j$172.4 & 194.4+$j$166.4 & 597.0+$j$296.2 \\
    Trust-region-reflective & 250.3+$j$172.4 & 194.5+$j$166.6 & 597.1+$j$296.4
    \end{tabular}
    \label{tab:line-parameters-estimation}
\end{table}

To evaluate whether these new values are a better estimate than datasheet information, we use a load flow to compute the voltage magnitude under the different impedance values and compare them against experimental measurements. The performance metric for node $k$ is the following
\begin{equation}
    \label{eq:voltage-error}
    \text{Error reduction}_k = \frac{V_{k,error}^{datasheet}-V_{k,error}^{estim}}{V_{k,error}^{datasheet}} \times 100\%,
\end{equation}
where $V_{k,error}^{datasheet}$ is the average error between the measured voltages $V_{k, t}^{meas}$ and the ones computed by load flow using the line parameters from the datasheet $V_{k,t}^{lf,datsasheet}$, given by
\begin{equation}
    V_{k,error}^{datasheet} = \frac{1}{T} \sum_{t=t_1}^{t^T}|V_{k, t}^{meas}-V_{k,t}^{lf,datasheet}| 
\end{equation}
Similarly, $V_{k,error}^{estim}$ is the average error between the measured voltages and the ones computed by load flow using the estimated line parameters $V_{k,t}^{lf,estim}$
\begin{equation}
    V_{k,error}^{estim} = \frac{1}{T} \sum_{t=t_1}^{t^T}|V_{k, t}^{meas}-V_{k,t}^{lf,estim}| 
\end{equation}
A perfect estimation of the line parameters would result in an error reduction of $100\%$.

\begin{figure}[]
    \centering
    \includegraphics[width=\linewidth]{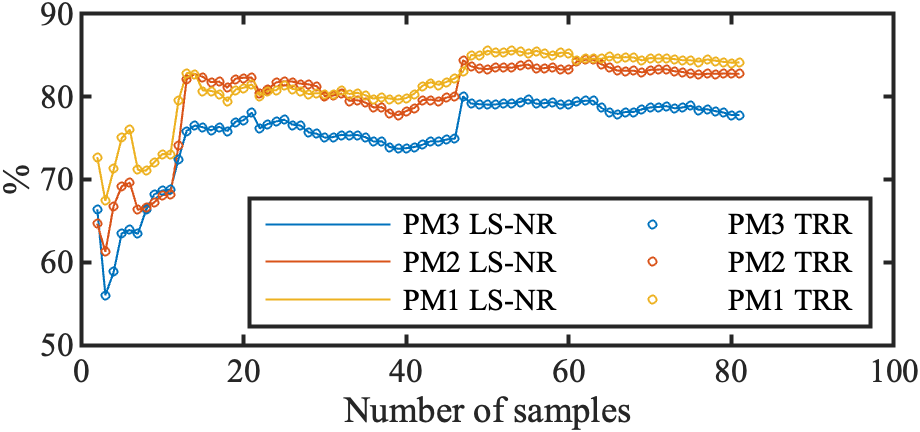}
    \caption{Voltage error reduction percentage as a function of the number of samples for Least-squares with Newton-Raphson (LS-NR) and Trust-region-reflective (TRR) for nodes PM1-PM3.}
    \label{fig:voltage-error}
\end{figure}

Fig.~\ref{fig:voltage-error} shows \eqref{eq:voltage-error} as a function of the number of samples used for LS, from 2 to 81, for both methods. The sample order is shuffled randomly. Both methods perform similarly and, when all data points are used, reduce the estimation error by up to 83\%. The maximum observed voltage error decreased from 0.02~pu (using datasheet parameters) to 0.008~pu if we use the updated parameters of Table~\ref{tab:line-parameters-estimation}.
Beyond approximately 50 data points, the improvement plateaus. Because the 81 data points cover the full range of the grid's capabilities, we can assume that the estimation of the line parameters is reliable. If fewer data points are used, the power injections should be uniformly distributed according to the grid's capabilities. The maximum current measured per phase was 7~A, much less than the sensor's nominal current (50~A); therefore, the line-parameters estimation is reliable even with non-negligible measurement errors. However, it was observed that less accurate sensors degraded the estimation performance.

\begin{figure}[]
    \centering
    \includegraphics[width=\linewidth]{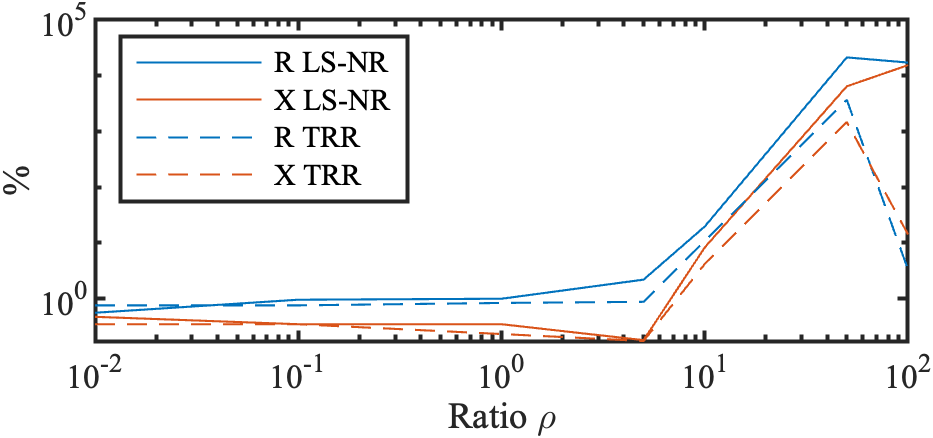}
    \caption{Maximum estimation error of the resistance R and the reactance X achieved by Least-squares with Newton-Raphson (LS-NR) and Trust-region-reflective (TRR) as a function of the ratio $\rho$ given by \eqref{eq:ratio-r}.}
    \label{fig:z_estimation_error}
\end{figure}

Next and finally, we study the convergence of the two methods for different initial estimates of the line parameters. The impedances estimated in Table~\ref{tab:line-parameters-estimation} are assumed to be the real ones, $Z_l^{real}$; the initial estimates are modified as
\begin{equation}
    \label{eq:ratio-r}
    Z_{l,0}^{estim} = \rho Z_l^{real}
\end{equation}
for $\rho \in [0.01,100]$. We compute the final percentage error in estimating the line impedances achieved by NR and TRR after they have converged. 
Fig.~\ref{fig:z_estimation_error} shows the maximum error over all lines achieved by the two algorithms, for R and X separately. For $\rho < 10$, the two methods converge to the true values with high accuracy ($< 1\%$), with TRR achieving slightly smaller values.
For larger $\rho$, convergence to the true parameters is not guaranteed. Nevertheless, the proposed algorithms perform reliably even without available datasheet values, provided a low initial estimate is used.

As a final note, TRR runs faster than NR, provided that we supply the analytical computation of the Jacobian. If we use all 81 available data points, TRR takes 0.05~s to converge if we provide the Jacobian, whereas it takes 9.2~s if we do not provide it. On the other hand, NR takes 1.9~s, even though it uses the analytical computation of the Jacobian. However, the difference might be due to our custom implementation of NR.

\section{Conclusion} \label{sec:concs}
We have proposed and experimentally validated various methods to estimate the series resistance and reactance of distribution grids' power lines. We examined two cases, one featuring phasor measurements and the other with RMS measurements, as in LV grids, smart meters might be the only source of information. In the first case, the system of power flow equations is fully determined, whereas in the second case, at least two measurements at different time instances are needed, provided that the two sets of injected powers differ adequately. We have also formulated a least-squares estimation problem in two ways, using Newton-Raphson and quadratic optimization, to estimate the line impedances in the presence of measurement errors and noise.

The methods were validated and compared both in simulations and experimentally. The simulations verified that two measurements at different time intervals are sufficient in the proposed setting (full observability) to correct the estimates of the line parameters when the initial estimates are not too dissimilar from the actual values. The experimental validation showed that the proposed least-squares approach improves the estimation of the line impedances compared to datasheet values, reducing the error in voltage estimation.
The corrected estimates could be helpful to improve grid-state estimation and load-flow accuracy.
Future research could generalize the methods to three-phase networks with unbalanced loads.


\bibliographystyle{IEEEtran}
\bibliography{IEEEabrv,main}

\end{document}